\documentclass[fleqn,10pt]{wlscirep}
\title{A novel procedure for the identification of chaos in complex biological systems}

\author[1,*]{D. Bazeia}
\author[2]{M.B.P.N. Pereira}
\author[2]{A.V. Brito}
\author[3]{B.F. de Oliveira}
\author[1]{J.G.G.S. Ramos}
\affil[1]{Departamento de F\'isica, Universidade Federal da Para\'iba, Jo\~ao Pessoa, Para\'iba, 58051-970, Brazil}
\affil[2]{Centro de Inform\'atica, Universidade Federal da Para\'\i ba, Jo\~ao Pessoa, Para\'\i ba, 58055-000, Brazil}
\affil[3]{Departamento de F\'isica, Universidade Estadual de Maring\'a, Maring\'a, Paran\'a, 87020-900, Brazil}

\affil[*]{dbazeia@gmail.com}


\begin{abstract}
We demonstrate the presence of chaos in stochastic simulations that are widely used to study biodiversity in nature. The investigation deals with a set of three distinct species that evolve according to the standard rules of mobility, reproduction and predation, with predation following the cyclic rules of the popular rock, paper and scissors game. The study uncovers the possibility to distinguish between time evolutions that start from slightly different initial states, guided by the Hamming distance which heuristically unveils the chaotic behavior. The finding opens up a quantitative approach that relates the correlation length to the average density of maxima of a typical species, and an ensemble of stochastic simulations is implemented to support the procedure. The main result of the work shows how a single and simple experimental realization that counts the density of maxima associated with the chaotic evolution of the species serves to infer its correlation length. We use the result to investigate others distinct complex systems, one dealing with a set of differential equations that can be used to model a diversity of natural and artificial chaotic systems, and another one, focusing on the ocean water level.
\end{abstract}
\begin{document}

\flushbottom
\maketitle
\thispagestyle{empty}

%%%%%%%%%%%%%%%%%%%%%%%%%
\section*{Introduction}

In his book on the origin of species, Darwin settled the foundation of evolutionary biology, foreseen the interplay between non-linearity and stochastic processes. Due to the unpredictability of the systems that evolve driven by natural selection rules, over the years several approaches to investigate the problem have been developed. The general procedure relies on eliminating irrelevant information to deal with the appropriate issue, as one sees, for instance, in the Brownian motion, the Langevin and the Fokker-Planck equations, and the Onsager, Uhlenbeck, and Chandrasekhar theories.  

Nowadays, despite the many ways to study complex systems, stochastic simulations have became a universal tool, and have been largely used to investigate collective behavior in nature. Based on a set of simple rules that are assessed randomly, the procedure has been employed in a diversity of scenarios to model and understand the experimental data. For a small set of investigations concerned mainly with biodiversity, see, e.g., the works\cite{1,2,3,4,5,6,7,8,9,10,11,12} and references therein. 

Without loss of generality, in this work we consider stochastic network simulations of the May and Leonard type \cite{5,7}, following two recent investigations, one describing how local dispersal may promote biodiversity in a real-life game \cite{8} and the other suggesting that population mobility may be central feature to describe real ecosystems \cite{10}. These studies develop simulations that engender cyclic competition, which is modeled as in the rock-paper-scissors game, controlled by the simple rules: paper wraps rock, rock crushes scissors and scissors cut paper \cite{6}. We consider the standard system with three distinct species $a$, $b$, and $c$, identified with the colors red, blue, and yellow, respectively, and one notes an interesting behavior of the stochastic simulations which is the pattern formation, indicating a subjacent law. Thus, the separation between the random and the dynamical and kinematical mechanisms that may perhaps contribute to the temporal evolution is of fundamental importance to its understanding.

The study focuses mainly on the presence of chaos, and we go on motivated by the fact that the mechanisms that control the evolution of the system are capable of imprinting the spatial patterns, with the pattern formation being sensible to subtle changes in the initial conditions, driven by the number of individuals in each species. The subject has been investigated before in many different contexts, with very interesting works being carried out on the chaotic behavior in biological systems and in other stochastic situations. Here we recall the works \cite{C1,C2,C3,C4,C5,C6,C7,C8,C9,O1,O2,O3,O4,O5,O6,O7,R1,R2,R3} that are closely related to the current investigation. In particular, in \cite{C1,C2,C3,C4,C5,C6,C7,C8,C9} the authors put a great deal of effort into the identification of chaos, but in the current work we suggest a new route, which is based on the Hamming distance concept \cite{13}, and on the counting of the density of maxima \cite{14} of the abundance of the species, which we explain below. Other issues that are considered in \cite{O1,O2,O3,O4,O5,O6,O7,R1,R2,R3} deal with generalizations of the standard three-species system to consider systems with four or more species, with the addition of other rules that may jeopardize biodiversity. Specific possibilities are studied in \cite{O3} with the addition of diffusion, which may block the spiral pattern formation due to mobility, and also in \cite{O5}, when one adds players that never change their strategy, regardless of the neighborhood. Moreover, in the recent work \cite{O7} the authors investigate how to preserve biodiversity despite the presence of a spatially heterogeneous environment. 

There are other studies on the cyclic dominance in evolutionary games, as one can find, for instance, in the recent review \cite{O6}. Some works investigate mechanisms that can be used to describe distinct kind of invasion rates between competing species, to act to jeopardize biodiversity. Specific examples can be found in the recent investigations \cite{R1,R2,R3}, which can be of direct interest to the current study, as we further comment in the Sec. Results.

We study the system with three distinct species and standard rules of evolution. The main motivation to deal within this well-known environment is to unveil the chaotic behavior incontestably. We move on encouraged by decades of studies in the physics of compound nuclei and nuclear resonances, and more recently by theories developed to investigate mesoscopic systems \cite{14}, in order to offer a positive answer to this issue. Specifically, we identify the presence of chaos in the stochastic evolution and use it to quantify correlation measurements through a single and simple experimental realization of the system. As we show below, instead of following the standard way to describe the chaotic behavior, we pave a new route, with the result relating the correlation length to the average density of maxima of the abundance or density of individuals of a typical species in the system. It is the main result of the work, and is due to the chaotic properties of the stochastic evolution, which we identify and measure in the current study. We also show that it is a general result and can be used in a diversity of situations, to infer the correlation length of systems that evolve in time chaotically. We first identify the finding, and then illustrate it with two distinct applications, one dealing with a set of first-order differential equations, and the other with the ocean water level.

%%%%%%%%%%%%%%%%%%%%%%%%%%%%%
\section*{Results}

\begin{figure}[t]
\begin{center}
{\includegraphics[width=12cm]{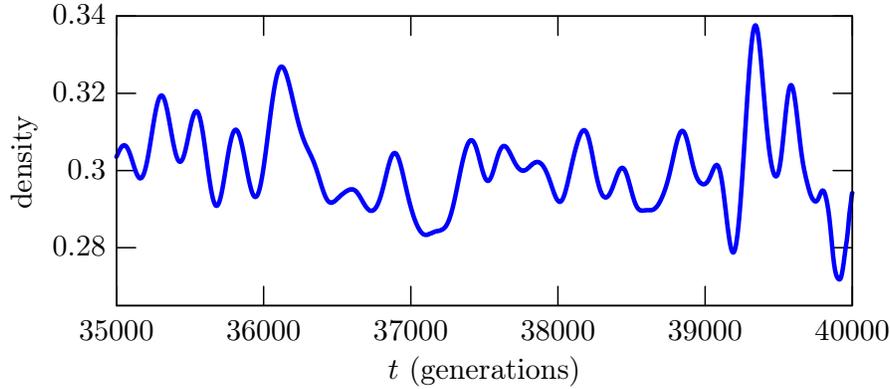}}
\caption{The blue B\'ezier curve shows the evolution of the abundance of a typical species in the interval in between 35000 and 40000 generations. The curve is built from the data corresponding to the blue species that appears in Fig.~\ref{fig4}, as described in Sec. Methods.}\label{fig1}
\end{center}
\end{figure}

We use stochastic simulations and follow standard procedure to investigate a system with three distinct species \cite{8,10}. The approach is implemented in Sec. Methods, and there one includes detailed investigation that allows to develop all the results that we describe in the current Section. In particular, one describes the abundance or density of individuals of all the three species. A typical result appears in Fig.~\ref{fig1}, where one uses the B\'ezier algorithm to display in blue the abundance of a given species. It shows a pattern that evolves in time oscillating around an average value, developing an apparently unpredictable number of maxima. As it is known, however, the cyclic behavior shown in the figure is in general required to generate stable evolution in a system described by a set of species that fight among themselves under specific rules. In a relatively short time scale, some species may tend to be suppressed, while others increase their abundance, but the dominance of a typical species over another one changes with the time, decreasing and inverting behavior, inducing an average value in a long time evolution. This cyclic equilibrium behavior is the central issue of the current work, and is used to identify the chaotic behavior even in a single and short time evolution of the system, as it is assured by the maximum entropy principle and the ergodic theorem.

\begin{figure}[t]
\begin{center}
\fbox{\includegraphics[width=10cm]{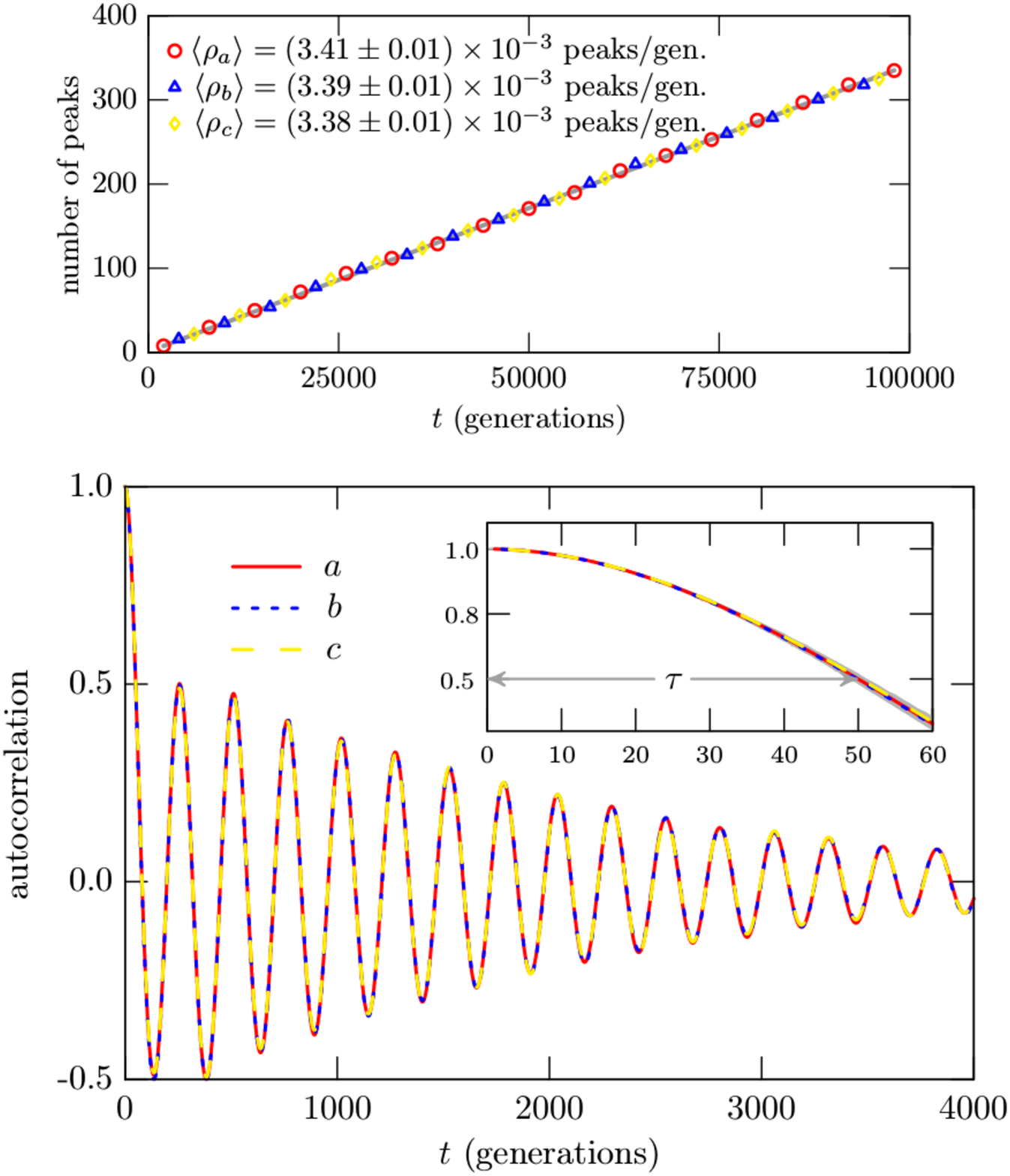}}
\caption{(Top) The counting of the number of peaks for the three species with the corresponding colors blue, red and yellow, with an almost invisible gray line depicting the fitting curve. (Bottom) The correlation function is displayed for the three species in a long time evolution, showing the general behavior. In the inset, it is depicted for a short time evolution, enough to show the correlation length. The shadow region in the inset displays the standard deviation. See Sec. Methods for details.}\label{fig2}
\end{center}
\end{figure}

The pattern displayed in Fig.~\ref{fig1} presents a number of maxima that can be related to the correlation length, due to the intrinsic chaotic behavior of the density of individuals. The study allows us to express the main result of the current work in the form 
\begin{equation}\label{result}
\tau=\frac1{6\left<\rho\right>}.
\end{equation}
It relates the correlation length $\tau$ to the average density of maxima $\left<\rho\right>$ of the abundance of a typical species which is illustrated in Fig.~\ref{fig1}. The counting of maxima is related to the tendency of the system to return to equilibrium, so the higher the density of maxima, the faster the tendency to approach equilibrium, although the system never relax to the state with the abundance fixed at the corresponding average value. 
The quantification that appears in Eq.~\eqref{result} is due to the chaotic pattern of the stochastic evolution, which we identify and measure using standard techniques. Before working this out, however, in this Fig.~\ref{fig1} one counts the number of maxima in the interval in between $35000$ and $40000$ generations to get the average density $\left<\rho\right>\approx0.0034$, compatible with the values indicated in Fig.~\ref{fig2} (Top). We then use \eqref{result} to get $\tau\approx49$, consistent with the value suggested in the inset in Fig.~\ref{fig2} (Bottom). The results displayed in Fig.~\ref{fig2} are obtained from an ensemble of stochastic simulations which we explain in Sec. Methods.

%%%%%%%%%%%%%%%%%%%%%%%%%%%%%%%%
%%%%%%%%%%%%%%%%%%%%%%%%%%%%

%%%%%%%%%%%%%%%%%%%%%%%%%
%%%%%%%%%%%%%%%%%%%%%%%

%%%%%%%%%%%%%%%%%%%%%%
\begin{figure}[t]
\begin{center}
\includegraphics[width=12cm]{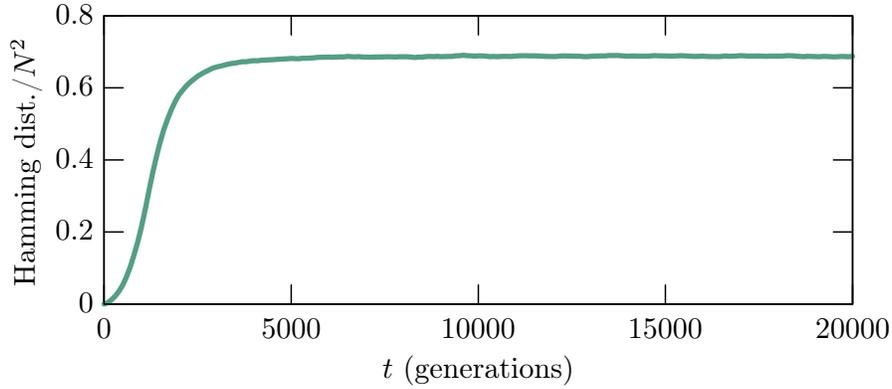}
\caption{The Hamming distance density is shown in green in a long time interval. It represents an average obtained from a set of 100 simulations, each one starting with a distinct initial state.}\label{fig3}
\end{center}
\end{figure}
%%%%%%%%%%%%%%%%%%%%%%%%%%%%%

%%%%%%%%%%%%%%%%%%%%%%%%%%%%%
\begin{figure}[t]
\begin{center}
\includegraphics[width=12cm]{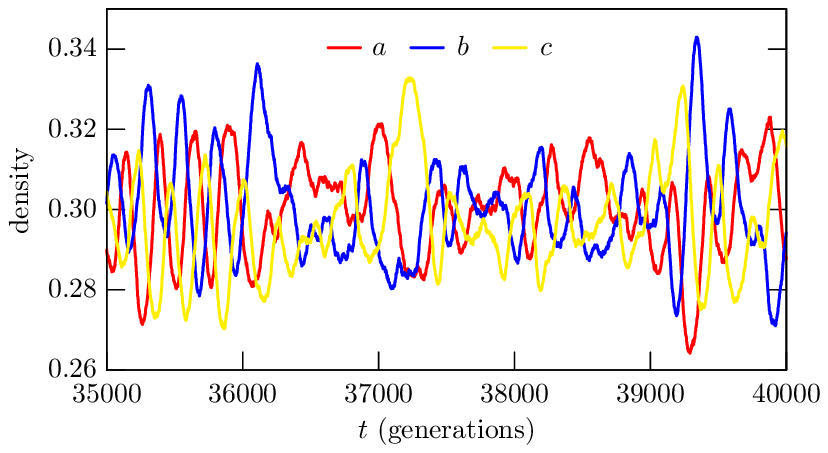}
\caption{Evolution of the abundance of the three species, depicted with the corresponding colors, in the long interval in between 35000 and 40000 generations. The data corresponding to the blue species is used to depict the B\'ezier curve in Fig.~\ref{fig1}.}\label{fig4}
\end{center}
\end{figure}
%%%%%%%%%%%%%%%%%%%%%%%%%%%%%%

The above expression \eqref{result} is of very practical use, because it allows us to infer the correlation length with a single and simple assessment to the time evolution of the chaotic variable under consideration, being it of natural or artificial origin, so it is of wide applicability. It follows from the identification that the stochastic simulation which is generically used in biodiversity engenders chaotic behavior. To focus on this, in Sec. Methods one describes the numerical procedure to be employed in this work. 
We then use it to assess the random processes engendered by the dynamical evolution. However, due to the impossibility to directly control the randomness of the simulations, the strategy elaborated follows a procedure which generates two distinct final states starting from initial states that are slightly different from each other.

Owing to quantify the chaotic behavior, we employ the Hamming distance \cite{13} to measure the difference between the two final states, which is a simple and appropriate manner to distinguish quantities such as vectors, matrices, etc, and can be illustrated with binary vectors. Since it counts the number of sites in the second vector that do not match with the corresponding sites of first one, the distance between $(0,0,0,1)$ and $(1,0,1,1)$ is two, for instance. We use the Hamming distance $H(t)$ to measure the difference between the two final states, which are now seen as two distinct $N\times N$ matrices. In Fig.~\ref{fig3} one displays the Hamming distance density $h(t)=H(t)/N^2$ with the solid green curve that represents an average obtained from a set of 100 simulations, each one starting with a distinct initial state. We recall that it is actually nontrivial to distinguish chaos from randomness, but here one has to emphasize that the new algorithm that we have implemented to account for the Hamming distance, which is described in Sec. Methods, provides the possibility to access the chaotic behavior that underlies the stochastic simulations very clearly.

The behavior that appears in Fig.~\ref{fig3} is a key result of the work. It shows that the Hamming distance density increases smoothly and then converges to a given value inside an interval with very narrow width. It is asymptotically stable and does not fill the full lattice. For an animated illustration on this, see the video at \url{https://youtu.be/nzLP-XcCvWc}. We have implemented other simulations, changing the values of the parameters $(m,r,p)$, the size of the lattice and the number of species in the system, and observed the same qualitative behavior, showing its universality.

As we have further noted, if one modifies the rules that control the time evolution, the profile of the Hamming distance density displayed in Fig.~\ref{fig3} may change drastically. For instance, if the modification of the rules does not jeopardize biodiversity, the Hamming distance density gets the universal behavior shown in Fig.~\ref{fig3}. However, if one follows the investigations considered in \cite{R1,R2,R3}, for instance, and modifies the rules in order to destroy biodiversity, the Hamming distance density starts increasing, oscillating in the interval $[0,1]$, but it ends up vanishing in the long run, if the two initial states evolve to become the same final state, or become unit, if the two final states are completely different from each other. The result shows that the asymptotic stability of the Hamming distance density is intrinsically connected with biodiversity. We are now investigating how the amplitude and width of the Hamming distance density behave for three, four and five species, as one increases the mobility to higher and higher values, reaching the region that jeopardizes biodiversity, as firstly shown in \cite{10} and further explored in the more recent works \cite{N1,N2}. We hope to offer a detailed investigation on this issue in the near future, emphasizing that the Hamming distance behavior is directly related with the evolution of the abundance of a typical species that is displayed in Fig.~\ref{fig1}, which was built from the data corresponding to the blue species that appears in Fig.~\ref{fig4}. The results depicted in Fig.~\ref{fig4} are further described in Sec. Methods, and show that the abundance of each one of the three species fluctuates around the same average value, as shown in red, blue and yellow.

Knowing that the stochastic processes engender chaotic behavior, it is of interest to further explore the above result. Here one should ask for the behavior of the system under perturbations, the presence of attractors, the Lyapunov spectrum etc. Instead of this standard route, however, we move forward inspired by the previous study \cite{14} and focus attention on counting the density of maxima of the time evolution of the abundance of the species. In this sense, one takes $l_i$ to describe the abundance of one of the three species $i=a$ (red), $b$ (blue), or $c$ (yellow), and considers the investigation developed at the end of the Sec. Methods to calculate the average density of maxima $\left<\rho_i \right>$. The main results are shown in Eqs.~\eqref{densidade1}--\eqref{p0l}, and can be used to write $\left<\rho_i \right>$ in the form 
\begin{equation}\label{rho}
\left< \rho_i\, \right>=\frac{1}{2 \pi} \sqrt{\frac{\left< l_i^{\prime\prime2} \right>}{\left< l_i^{\prime2} \right>}}\,.
\end{equation}
This is another key result of the work, and offers a new quantity that arises from the chaotic behavior imprinted in the stochastic simulations.
To verify its validity, we implemented a procedure to create an ensemble of events to obtain the correlation function, which is used to calculate the correlation length. As it is known, the correlation length $\tau$ is extracted from the correlation function as the width at half height, so we use the fitting function $C(t)=\cos(\kappa\,t)$ to write $\kappa\,\tau=\pi/3$, and from Eqs.~\eqref{rho} and \eqref{med} one gets to
\begin{equation}
\left< \rho\, \right>=\frac{\kappa}{2 \pi}.
\end{equation}
This allows to relate the density of maxima to the correlation length in the form $\tau=1/(6\left<\rho\right>)$, as we have anticipated in Eq.~\eqref{result}. 

The result offers a way to measure the correlation length from a single experimental assessment to the system under investigation, calculating the average density of maxima, as it is illustrated in Fig.~\ref{fig1}. We note that the finding nicely connects the top and bottom panels in Fig.~\ref{fig2}, since the correlation length which one reads from the inset in the bottom panel in Fig.~\ref{fig2} can be obtained from the value of the density of maxima which one reads from the top panel in Fig.~\ref{fig2}, via Eq.~\eqref{result}.

\subsection*{Applications}

\bigskip

\begin{figure}[ht]
\begin{center}
\fbox{\includegraphics[width=16cm]{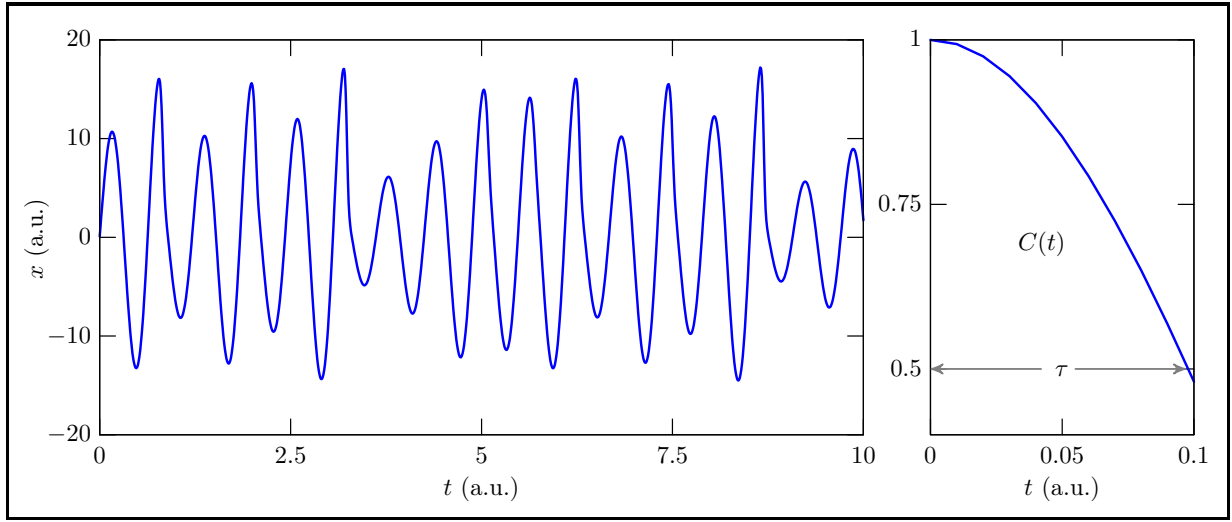}}
\caption{The blue curve in the left panel describes the $x(t)$ evolution that follows from the set of equations \eqref{ros}. The right panel displays the corresponding correlation function.}\label{fig5}
\end{center}
\end{figure}

The investigation described above shows that the result in Eq.~\eqref{result} is valid in a broader context, and can be used to investigate other systems that develop chaotic behavior. To illustrate this issue, in Fig.~\ref{fig5} one displays how $x$ evolves in time for the deterministic system of first-order differential equations
\begin{equation}\label{ros}
\frac{dx}{dt}=-y-z, \;\;\;\;\;\frac{dy}{dt}=x+0.15\, y;\;\;\;\;\;\frac{dz}{dt}=0.2-10\, z+x\, z.
\end{equation}
They are the R\"ossler equations \cite{15}, and can be used to model a diversity of natural and artificial chaotic systems. We see from the left panel in Fig.~\ref{fig5}, where the behavior of $x$ is shown, that the number of maxima is 17, so the correlation length gives $\tau=0.097$. This is an excellent result, since using the fitting function for the correlation function displayed in the right panel in Fig.~\ref{fig5}, one gets to $\tau=0.098$.

\begin{figure}[ht]
\begin{center}
\fbox{\includegraphics[width=16cm]{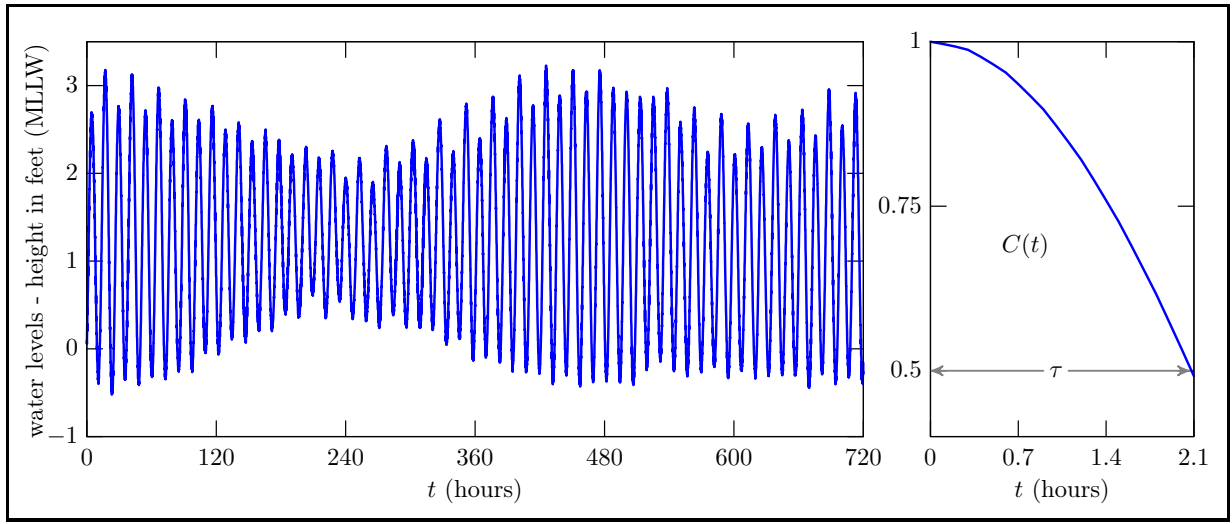}}
\caption{The ocean water level for the station ID 1770000, Pago Pago, American Samoa, is shown in the left panel for the period from 01/Aug/2016 - 00h00 GMT to 30/Aug/2016 - 23h59 GMT. The right panel displays the corresponding correlation function.}\label{fig6}
\end{center}
\end{figure}

We move on to examine another system of distinct nature that also evolves chaotically. We focus attention on the ocean water level, with data available from the National Oceanic and Atmospheric Administration. This brings about another illustration, which appears in Fig.~\ref{fig6}, displaying in blue the observed water level for the station ID 1770000, Pago Pago, American Samoa. See {\url{http://tidesandcurrents.noaa.gov/waterlevels.html?id=1770000}}. In the right panel in Fig.~\ref{fig6} one shows the correlation function, visually indicating that the correlation length is around $2$ hours. In fact, if one uses the fitting curve it gives $\tau= 2.0748$, and if one counts the maxima that appear in the left panel in Fig.~\ref{fig6} and then uses the result in Eq.~\eqref{result}, one gets to the value $\tau=2.0689$.

%%%%%%%%%%%%%%%%%%%%%%%%%%%%%%
%%%%%%%%%%%%%%%%%%%%%%%%%%%%%%%%
%%%%%%%%%%%%%%%%%%%%%%%%%%%%%%%%%

%%%%%%%%%%%%%%%%%%%%%%%%%%%%
\section*{Discussion}

In this work we studied the presence of chaos in a system with three distinct species, that evolve under the standard rules of mobility, predation and reproduction. We first used the Hamming distance density to heuristically characterize the chaotic behavior embedded in the stochastic evolution of the system. The typical result is shown in Fig.~\ref{fig3}, and is obtained from the algorithm developed in Sec. Methods, which effectively separates the chaotic behavior from the randomness of the stochastic simulations. The qualitative behavior displayed in Fig.~\ref{fig3} suggests that we investigate how the amplitude and width of the Hamming distance density behave for three, four and five species, as one increases the mobility to higher and higher values, reaching the region that jeopardizes biodiversity. Being a quantity that unveil the chaotic behavior, the Hamming distance density may be used as an order parameter to investigate chaos in complex systems. We will report on this issue in the near future.

The finding motivated us to developed a stronger result, unveiling a quantitative approach to the correlation length from a single and simple experimental measurement of the average density of maxima that appear in the abundance of a typical species during the dynamical evolution of the system. The connection between the density of maxima and the correlation length was formulated under the principle of maxima entropy, using an ensemble of events that we developed numerically in Sec. Methods. Since the stochastic simulations here considered are universally used to describe biodiversity in nature, we are then left with the suggestion that chaos is imprinted in biodiversity. 
 
The study unveiled a simple route to obtain the correlation length associated to a given chaotic time evolution, and the simplicity of the procedure makes the result especially relevant to investigate complex chaotic systems. Examples of applications of this result abound, and here we have illustrated its applicability with the deterministic system described by the set of first-order differential equations \eqref{ros}, which are known as the R\"ossler equations \cite{15} and engender the behavior displayed in Fig.~\ref{fig5}. Also, we have considered another well distinct complex system, dealing with the ocean water level data that is displayed in Fig.~\ref{fig6}. 

The main result of this work indicates that a single and simple experimental investigation of the time evolution of a given chaotic quantity allows estimating the correlation length of the corresponding quantity with a very high confidence level. As we have illustrated with some explicit examples, it can be widely used to study complex processes, as the ones that appear in Geology, Meteorology, Evolutionary Biology, and in many other areas of nonlinear science. We hope that the current study may help us to better understand biodiversity and its associated chaotic behavior, inspiring further research on similar issues. In particular, the simplicity of the result suggests that the ergodic hypothesis and the maximization of the entropy are valid for complex biological systems and may serve for other studies on the thermodynamics and statistical properties of such systems.

%%%%%%%%%%%%%%%%%%%%%%%%%%%%%%
\section*{Methods}

In this work we developed stochastic simulations considering a standard system which is widely used to investigate biodiversity in nature. We review the procedure recalling that the simulations are implemented considering the system with three distinct species $a$, $b$, and $c$, identified with the colors red, blue, and yellow, respectively. They evolve according to the three basic rules: mobility $(m)$, reproduction $(r)$, and predation $(p)$. The evolution is carried out standardly, taking a square lattice of size $N\times N$, in which the three species and the empty sites (which is identified by $e$, with the color white) are equally but randomly distributed in the lattice, such that the quantity of individuals of each species (including the empty sites) is $L=N^2/4$, at the initial state. We deal with local dispersion and appropriate mobility. A given site is considered active if occupied by an individual of one of the three active species, and it may interact with one of its eight nearest neighbors, the Moore neighborhood. 
%%%%%%%%%%%%%%%%%%%%%%
\begin{figure}[ht]
\begin{center}
\fbox{\includegraphics[width=16cm]{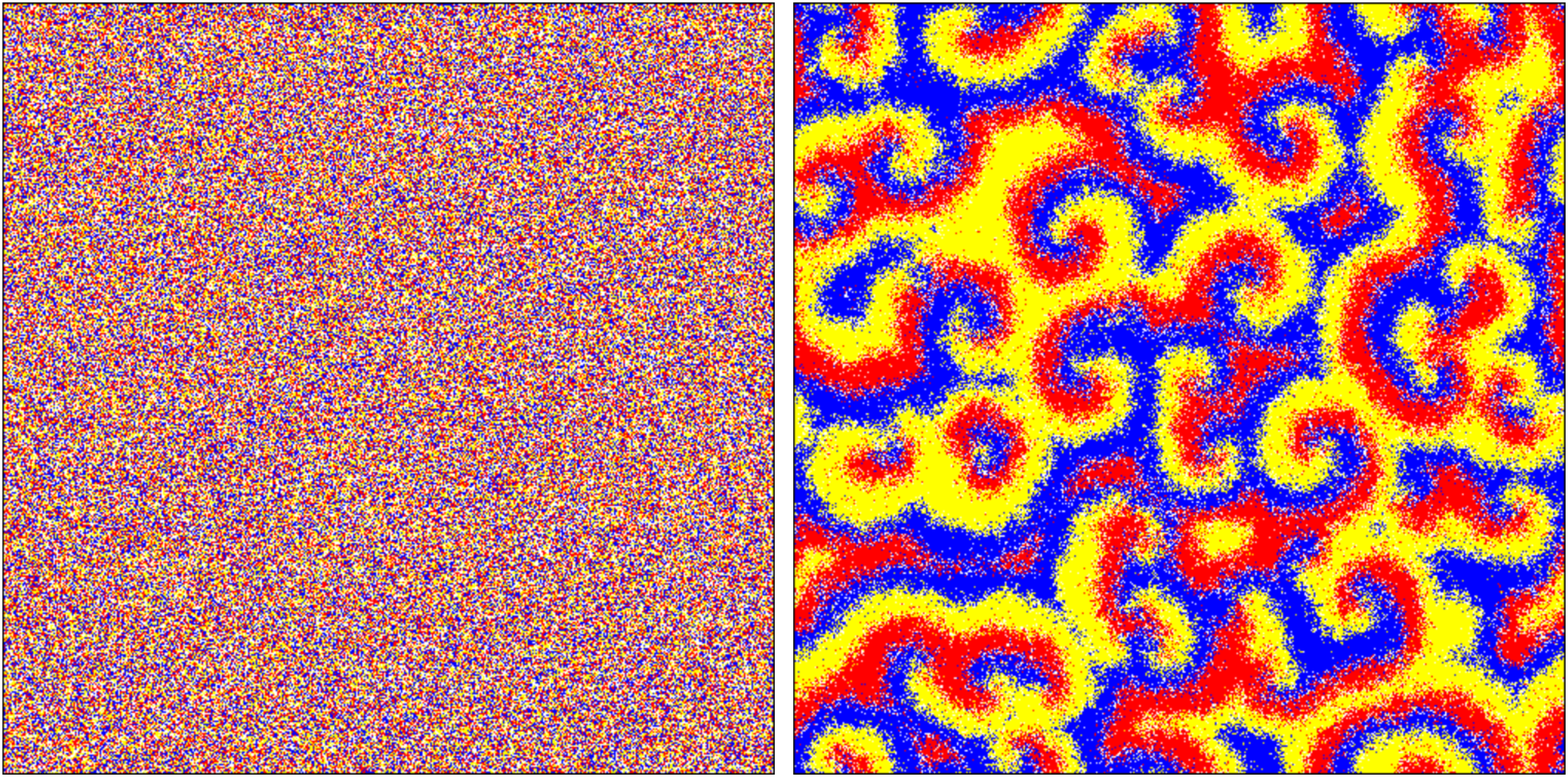}}
\caption{The $500 \times 500$ square lattice is shown at the initial time (left), and after $2000$ generations (right), illustrating the formation of spatial patterns.}\label{fig7}
\end{center}
\end{figure}
%%%%%%%%%%%%%%%%%%%%%%%%%

As usual, the numerical simulations are performed with periodic boundary conditions. If $i$ stands for $a,b,$ or $c$, and $\alpha$ for $a,b,c$, or $e$, mobility and reproduction are represented by $i\,\alpha\to \alpha\,i$ and $i\,e\to i\,i$, respectively. The other rule, predation, follow the rock-paper-scissors game, that is, $a\,b\to a\,e$, $b\,c\to b\,e$, and $c\,a\to c\,e$. In this work we take $m=0.5$, $r=0.25$, and $p=0.25$, for mobility, reproduction and predation, respectively, for all the three species. We have considered other possibilities, with $m=0.25$, $r=0.25$ and $p=0.5$, for instance, and no qualitative modification in the results was found. The dynamical process starts with a random assess to the square lattice, followed by a random selection of one of the three rules and by a random choice of one of the eight neighbors. It is then checked if the assessed site is empty or not: if it is empty, one returns to the lattice to simulate another assess to it; if it contains an individual of one of the three species, one takes the selected rule and uses it with the selected neighbor. To measure the time evolution we use generation, which is the time spent to access the lattice $N^2$ times. 

To prepare the initial state, one randomly chooses one among the three species $a$, $b$ and $c$, and the empty site $e$ with the same probability, and distributes it in the square lattice. The procedure is repeated $N^2$ times, evenly filling all the sites in the square lattice. A typical initial state is illustrated in the left panel in Fig.~\ref{fig7}. One uses it to run the stochastic simulations and get to the final configuration which is displayed in the right panel of Fig.~\ref{fig7}. There one sees that the system evolves forming a specific pattern, in which the species organize themselves in spatial portions of the lattice. This is known in the literature, and has been explored in a diversity of contexts, to study diversity and other related behaviors.

In order to unveil the chaotic behavior, we use the stochastic simulations to generate a new algorithm. The procedure goes as follows: one first builds a initial state as done in the left panel in Fig.~\ref{fig7}, and makes a copy of it. One uses this initial state to run the simulation to get to the final state, which is then saved. The key point here is that during the time evolution a new file is created, in which one saves every single step used to run it. One then takes the copy of the initial state and randomly selects a lattice site and modifies its content. This new state has the tiniest difference, since among the many lattice sites it has a single site which is different from the initial state already used to evolve in time. With this new initial state, one runs the same simulation already considered, evolving it according to the very same rules, in the same order and pace, as they appear in the saved file. The procedure leads to another final state, which is also saved.

\begin{figure}[ht]
\begin{center}
\fbox{\includegraphics[width=16cm]{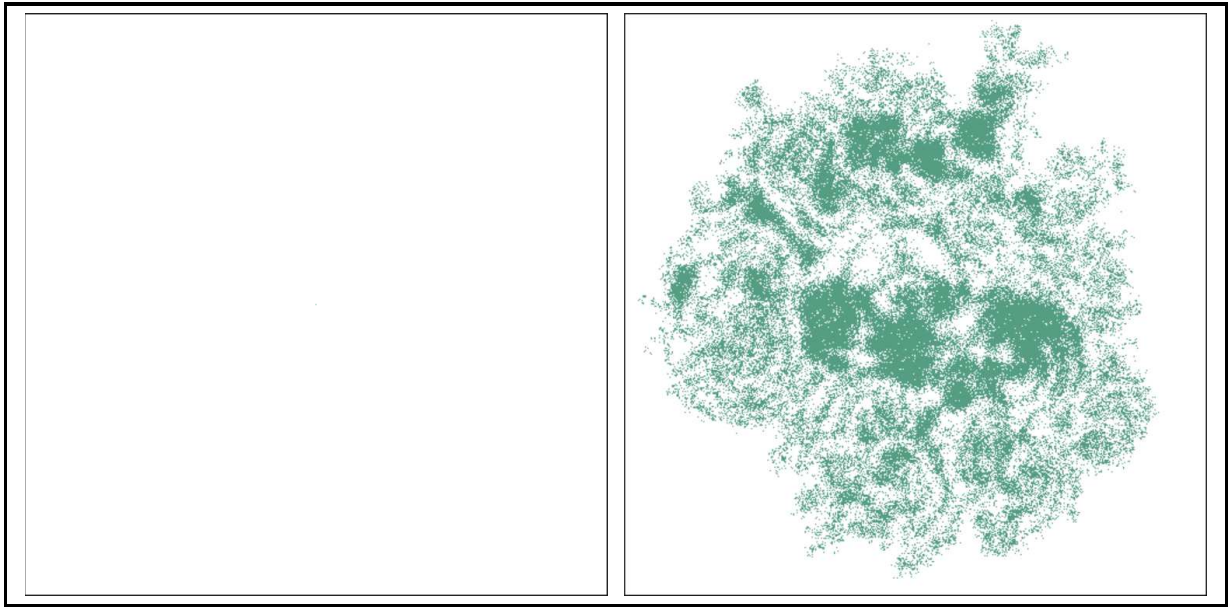}}
\caption{An illustration of the Hamming distance at the initial time (left), showing an almost invisible green dot at the center of the lattice, and after 1000 generations (right), indicating how the two states separate from each other.}\label{fig8}
\end{center}
\end{figure}

The two final states are different, but the difference has nothing to do with the randomness of the stochastic simulations, being due to the tiniest modification introduced in the second initial state. To infer the presence of chaos, one has to measure the difference between the two final states, generated with the same stochastic rules. Toward this goal, we have developed an algorithm which displays in light green the lattice sites that differ in the two configurations, leaving the other sites empty. In Fig.~\ref{fig8} two snapshots are displayed, one at the initial time in the left panel, with a single almost invisible light green site (which is depicted at the grid center) marking the tiniest difference between the two initial states. The other snapshot is in the right panel, and it shows all the light green sites that mark the difference between the two final states, after implementing the simulations for a long time. The two snapshots show the impact that the subtle modification introduced in the initial state causes in the evolution of the system. The result suggests that the time evolution that appears in the stochastic simulation engenders chaotic behavior. The procedure allowed to introduce the Hamming distance density which we discussed in Sec. Results and displayed in Fig.~\ref{fig3}.

%%%%%%%%%%%%%%%%%%%%%%%%%%%%
%%%%%%%%%%%%%%%%%%%%%%%%%

To further explore the stochastic network simulations, one investigates how the quantities of individuals $L_i\;( i=a,b,c)$ vary as the system evolves in time. In fact, one uses the densities $l_i=L_i/N^2$ and the results are depicted in Fig.~\ref{fig4}, showing that the abundance of each one of the three species fluctuates around the same average value, as shown in red, blue and yellow. They do not add to unit because of the number of empty sites, that fluctuates around a lower value. This happens because the species move, reproduce and predate evenly, while the empty sites enter the game passively. We have run other simulations, starting from an initial state with no empty site, and ended up with final states with the species having abundance similar to the ones displayed in Fig.~\ref{fig4}. This shows that the addition or not of empty sites in the initial state does not modify the abundance behavior for long time evolutions. The abundance behavior that appear in Fig.~\ref{fig4} show a substructure which is not present in the curve depicted in Fig.~\ref{fig1}, due to the B\'ezier algorithmic tool used to smooth its behavior. This is a legitimate procedure, since the substructure that appears in the abundance is due to the intrinsic discreteness of the stochastic approach, having nothing to do with the chaotic behavior present in the process under investigation. The B\'ezier algorithm that we implement in this work uses as the control points the set of points available from the data, in the time interval used to count the number of maxima.      

As can be seen from Fig.~\ref{fig4}, the abundance $l_i$ evolves in time and fluctuates to produce local maximum in the interval $[t,t+\delta t]$, for sufficiently small $\delta t$, so one has $l_i^\prime(t)>0,$ and $l_i^\prime(t+\delta t)<0,$ where prime stands for the time derivative, such that $-l_i''(t) \delta t\!>\! l_i'(t)>0$. The joint probability $P(l_i',l_i'')$ can be used to calculate the average density of maxima $\left< \rho_i\,\right>$ through the simple route: the probability to find a maximum in the interval $[ t,t + \delta t ]$ is proportional to the integral spanning the region defined above, such that
\begin{eqnarray}
\left< \rho_i\, \right>\equiv \frac{1}{\delta t}\! \int_{\!-\infty}^{0} \!\!\!\!d l_i^{\prime\prime}\!\! \int_{0}^{-l_i^{\prime\prime}\delta t}\!\! \!\!\!dl_i^\prime\, P(l_i^\prime,l_i^{\prime\prime})= -\!\! \int_{\!-\infty}^{0} \!\!\!\!dl_i^{\prime\prime}\, l_i^{\prime\prime}\, P(0,l_i^{\prime\prime}).\;\; \label{densidade1}
\end{eqnarray}
The fact that the statistical properties of the mean number of individuals of a given species are invariant under time translations indicates that both $l_i^\prime$ and $l_i^{\prime\prime}$ have vanishing average values. Moreover, the properties of $P(l_i^\prime, l_i^{\prime\prime})$ can be obtained from the smallest momenta
of $l_i^\prime$ and $l_i^{\prime\prime}$, and the variances of $P(l_i^\prime,l_i^{\prime\prime})$ are directly related to the correlation function
\begin{equation}
C_i(\delta t)= \left< l_i (t+\delta t)l_i (t) \right>.
\end{equation}
We can then obtain the several momenta, in particular
\begin{equation}\label{med}
\left< l_i^{\prime 2} \right> = - \frac{d^2 C_i(\delta t)}{d(\delta t)^2} \bigg|_{\delta t=0}\; ;\;\;\;
\left< l_i^{\prime\prime 2} \right> =  \frac{d^4 C_i(\delta t)}{d(\delta t)^4} \bigg|_{\delta t=0}.
\end{equation}

The principle of maximum entropy can be used to construct the joint probability distribution for $l_i$ and its derivatives from the previous equations. After implementing the algebraic calculations, integration on $l_i$ leads to $P(l_i',l_i'')$ which gives
\begin{equation}\label{p0l}
P(0,l_i^{\prime\prime})=\frac{1}{2 \pi} \frac{1}{\sqrt{\left< l_i^{\prime2} \right>\left< l_i^{\prime\prime2} \right>}}\exp\left( -\frac{1}{2} \frac{l_i^{\prime\prime2}}{\left< l_i^{\prime\prime2} \right>} \right).
\end{equation}
The above expressions can be used to write the density of maxima in terms of the correlation function, as we showed before in Sec. Results.
One then implement a numerical investigation to obtain an ensemble of events, taking an initial state and running the simulation for a very long time. This is a single event, and it is repeated many times, to get to the ensemble of events, each event being built as the first one, running the simulation from a random initial state within the same time interval. The ensemble is then used to count the number of peaks in $l_i\, (i=a,b,c)$ in the simulation. They are depicted in the top panel in Fig.~\ref{fig2}, for each one of the three species, and there the values of the average density of maxima are also shown.

The procedure is also used to get the correlation function, which is depicted in the bottom panel in Fig.~\ref{fig2} for all the three species in a long time evolution, showing the expected general behavior, with the periodicity reflecting the periodic boundary conditions in the lattice. In the inset we depict the correlation function for a time evolution enough to display the correlation length. One then uses the short time evolution shown in the inset of the bottom panel in Fig.~\ref{fig2} to search for the best fit curve, getting $C(t) = \cos(\kappa\, t),$ where $\kappa=0.0212$, with error lower then $5\%$, obtained within the least square approach. We emphasize that the natural decaying process present in the correlation function is not of practical significance to calculate the correlation length, since it can be obtained from the short time evolution displayed in the inset in the bottom panel in Fig.~\ref{fig2}.

\section*{Acknowledgements}

This work was partially financed by the CNPq Grants 455931/2014-3, 306614/2014-6, 308241/2013-4 and 479960/2013-5.

\section*{Author contributions statement}

M.B.P.N.P., A.V.B. and J.G.G.S.R. organized the numerical procedure. M.B.P.N.P., A.V.B. and B.F.O. implemented the stochastic simulations. D.B., B.F.O. and J.G.G.S.R. analysed the results and organized the figures. D.B. and J.G.G.S.R. wrote the manuscript. All authors reviewed and approved the work.

\section*{Additional information}

The authors declare no competing financial interests.


\begin{thebibliography}{10}
\expandafter\ifx\csname url\endcsname\relax
  \def\url#1{\texttt{#1}}\fi
\expandafter\ifx\csname urlprefix\endcsname\relax\def\urlprefix{URL }\fi
\expandafter\ifx\csname doiprefix\endcsname\relax\def\doiprefix{DOI }\fi
\providecommand{\bibinfo}[2]{#2}
\providecommand{\eprint}[2][]{\url{#2}}

\bibitem{1}
\bibinfo{editor}{David~Tilman, P.~K.} (ed.) \emph{\bibinfo{title}{Spatial
  Ecology: The Role of Space in Population Dynamics and Interspecific
  Interactions}} (\bibinfo{publisher}{Princeton University Press},
  \bibinfo{year}{1997}).

\bibitem{2}
\bibinfo{editor}{Ulf~Dieckmann, J. A. J.~M., Richard~Law} (ed.)
  \emph{\bibinfo{title}{The Geometry of Ecological Interactions: Simplifying
  Spatial Complexity (Cambridge Studies in Adaptive Dynamics)}}
  (\bibinfo{publisher}{Cambridge University Press}, \bibinfo{year}{2005}).

\bibitem{3}
\bibinfo{author}{Nowak, M.~A.}
\newblock \emph{\bibinfo{title}{Evolutionary Dynamics: Exploring the Equations
  of Life}} (\bibinfo{publisher}{Belknap Press}, \bibinfo{year}{2006}).

\bibitem{4}
\bibinfo{author}{Pinsky, M.~A.} \& \bibinfo{author}{Karlin, S.}
\newblock \emph{\bibinfo{title}{An Introduction to Stochastic Modeling}}
  (\bibinfo{publisher}{Academic Press}, \bibinfo{year}{2010}),
  \bibinfo{edition}{fourth edition} edn.

\bibitem{5}
\bibinfo{author}{May, R.~M.} \& \bibinfo{author}{Leonard, W.~J.}
\newblock \bibinfo{title}{Nonlinear aspects of competition between three
  species}.
\newblock \emph{\bibinfo{journal}{{SIAM} Journal on Applied Mathematics}}
  \textbf{\bibinfo{volume}{29}}, \bibinfo{pages}{243--253}
  (\bibinfo{year}{1975}).

\bibitem{6}
\bibinfo{author}{Sinervo, B.} \& \bibinfo{author}{Lively, C.~M.}
\newblock \bibinfo{title}{The rock-paper-scissors game and the evolution of
  alternative male strategies}.
\newblock \emph{\bibinfo{journal}{Nature}} \textbf{\bibinfo{volume}{380}},
  \bibinfo{pages}{240--243} (\bibinfo{year}{1996}).

\bibitem{7}
\bibinfo{author}{Durrett, R.} \& \bibinfo{author}{Levin, S.}
\newblock \bibinfo{title}{Spatial aspects of interspecific competition}.
\newblock \emph{\bibinfo{journal}{Theoretical Population Biology}}
  \textbf{\bibinfo{volume}{53}}, \bibinfo{pages}{30--43}
  (\bibinfo{year}{1998}).

\bibitem{8}
\bibinfo{author}{Kerr, B.}, \bibinfo{author}{Riley, M.~A.},
  \bibinfo{author}{Feldman, M.~W.} \& \bibinfo{author}{Bohannan, B. J.~M.}
\newblock \bibinfo{title}{Local dispersal promotes biodiversity in a real-life
  game of rock-paper-scissors}.
\newblock \emph{\bibinfo{journal}{Nature}} \textbf{\bibinfo{volume}{418}},
  \bibinfo{pages}{171--174} (\bibinfo{year}{2002}).

\bibitem{9}
\bibinfo{author}{Kirkup, B.~C.} \& \bibinfo{author}{Riley, M.~A.}
\newblock \bibinfo{title}{Antibiotic-mediated antagonism leads to a bacterial
  game of rock-paper-scissors in vivo}.
\newblock \emph{\bibinfo{journal}{Nature}} \textbf{\bibinfo{volume}{428}},
  \bibinfo{pages}{412--414} (\bibinfo{year}{2004}).

\bibitem{10}
\bibinfo{author}{Reichenbach, T.}, \bibinfo{author}{Mobilia, M.} \&
  \bibinfo{author}{Frey, E.}
\newblock \bibinfo{title}{Mobility promotes and jeopardizes biodiversity in
  rock-paper-scissors games}.
\newblock \emph{\bibinfo{journal}{Nature}} \textbf{\bibinfo{volume}{448}},
  \bibinfo{pages}{1046--1049} (\bibinfo{year}{2007}).

\bibitem{11}
\bibinfo{author}{Szab{\'{o}}, G.} \& \bibinfo{author}{F{\'{a}}th, G.}
\newblock \bibinfo{title}{Evolutionary games on graphs}.
\newblock \emph{\bibinfo{journal}{Physics Reports}}
  \textbf{\bibinfo{volume}{446}}, \bibinfo{pages}{97--216}
  (\bibinfo{year}{2007}).

\bibitem{12}
\bibinfo{author}{Claussen, J.~C.} \& \bibinfo{author}{Traulsen, A.}
\newblock \bibinfo{title}{Cyclic dominance and biodiversity in well-mixed
  populations}.
\newblock \emph{\bibinfo{journal}{Phys. Rev. Lett.}}
  \textbf{\bibinfo{volume}{100}}, \bibinfo{pages}{058104}
  (\bibinfo{year}{2008}).

\bibitem{C1}
\bibinfo{author}{Nowak, M.~A.} \& \bibinfo{author}{May, R.~M.}
\newblock \bibinfo{title}{Evolutionary games and spatial chaos}.
\newblock \emph{\bibinfo{journal}{Nature}} \textbf{\bibinfo{volume}{359}},
  \bibinfo{pages}{826--829} (\bibinfo{year}{1992}).

\bibitem{C2}
\bibinfo{author}{Mitchell, M.}, \bibinfo{author}{Hraber, P.~T.} \&
  \bibinfo{author}{Crutchfield, J.~P.}
\newblock \bibinfo{title}{Revisiting the edge of chaos: Evolving cellular
  automata to perform computations}.
\newblock \emph{\bibinfo{journal}{Complex Systems}}
  \textbf{\bibinfo{volume}{7}}, \bibinfo{pages}{89--130}
  (\bibinfo{year}{1993}).

\bibitem{C3}
\bibinfo{author}{Mitchell, M.}, \bibinfo{author}{Crutchfield, J.~P.} \&
  \bibinfo{author}{Hraber, P.~T.}
\newblock \bibinfo{title}{Evolving cellular automata to perform computations:
  mechanisms and impediments}.
\newblock \emph{\bibinfo{journal}{Physica D: Nonlinear Phenomena}}
  \textbf{\bibinfo{volume}{75}}, \bibinfo{pages}{361--391}
  (\bibinfo{year}{1994}).


\bibitem{C4}
\bibinfo{author}{Yang, C.~B.}, \bibinfo{author}{Cai, X.} \&
  \bibinfo{author}{Zhou, Z.~M.}
\newblock \bibinfo{title}{Spatial-temporal correlations in the process to
  self-organized criticality}.
\newblock \emph{\bibinfo{journal}{Phys. Rev. E}}
  \textbf{\bibinfo{volume}{61}}, \bibinfo{pages}{7243} (\bibinfo{year}{2000}).

\bibitem{C5}
\bibinfo{author}{Sato, Y.}, \bibinfo{author}{Akiyama, E.} \&
  \bibinfo{author}{Farmer, J.~D.}
\newblock \bibinfo{title}{Chaos in learning a simple two-person game}.
\newblock \emph{\bibinfo{journal}{Proceedings of the National Academy of
  Sciences}} \textbf{\bibinfo{volume}{99}}, \bibinfo{pages}{4748--4751}
  (\bibinfo{year}{2002}).

\bibitem{C6}
\bibinfo{author}{K{\'{a}}rolyi, G.}, \bibinfo{author}{Neufeld, Z.} \&
  \bibinfo{author}{Scheuring, I.}
\newblock \bibinfo{title}{Rock-scissors-paper game in a chaotic flow: The
  effect of dispersion on the cyclic competition of microorganisms}.
\newblock \emph{\bibinfo{journal}{Journal of Theoretical Biology}}
  \textbf{\bibinfo{volume}{236}}, \bibinfo{pages}{12--20}
  (\bibinfo{year}{2005}).

\bibitem{C7}
\bibinfo{author}{Gosak, M.}, \bibinfo{author}{Marhl, M.} \&
  \bibinfo{author}{Perc, M.}
\newblock \bibinfo{title}{Chaos between stochasticity and periodicity in the
  prisoner's dilemma game}.
\newblock \emph{\bibinfo{journal}{International Journal of Bifurcation and
  Chaos}} \textbf{\bibinfo{volume}{18}}, \bibinfo{pages}{869--875}
  (\bibinfo{year}{2008}).

\bibitem{C8}
\bibinfo{author}{Gosak, M.}, \bibinfo{author}{Marhl, M.} \&
  \bibinfo{author}{Perc, M.}
\newblock \bibinfo{title}{Chaos out of internal noise in the collective
  dynamics of diffusively coupled cells}.
\newblock \emph{\bibinfo{journal}{The European Physical Journal B}}
  \textbf{\bibinfo{volume}{62}}, \bibinfo{pages}{171--177}
  (\bibinfo{year}{2008}).

\bibitem{C9}
\bibinfo{author}{Nicolis, S.~C.} \emph{et~al.}
\newblock \bibinfo{title}{Foraging at the edge of chaos: Internal clock versus
  external forcing}.
\newblock \emph{\bibinfo{journal}{Phys. Rev. Lett.}}
  \textbf{\bibinfo{volume}{110}}, \bibinfo{pages}{268104}
  (\bibinfo{year}{2013}).


\bibitem{O1}
\bibinfo{author}{Szab{\'{o}}, G.}, \bibinfo{author}{Szolnoki, A.} \&
  \bibinfo{author}{Izs{\'{a}}k, R.}
\newblock \bibinfo{title}{Rock-scissors-paper game on regular small-world
  networks}.
\newblock \emph{\bibinfo{journal}{Journal of Physics A: Mathematical and
  General}} \textbf{\bibinfo{volume}{37}}, \bibinfo{pages}{2599--2609}
  (\bibinfo{year}{2004}).

\bibitem{O2}
\bibinfo{author}{Szolnoki, A.} \& \bibinfo{author}{Szab{\'{o}}, G.}
\newblock \bibinfo{title}{Phase transitions for rock-scissors-paper game on
  different networks}.
\newblock \emph{\bibinfo{journal}{Phys. Rev. E}}
  \textbf{\bibinfo{volume}{70}}, \bibinfo{pages}{037102}
  (\bibinfo{year}{2004}).


\bibitem{O3}
\bibinfo{author}{Peltom\"{a}ki, M.} \& \bibinfo{author}{Alava, M.}
\newblock \bibinfo{title}{Three- and four-state rock-paper-scissors games with
  diffusion}.
\newblock \emph{\bibinfo{journal}{Phys. Rev. E}}
  \textbf{\bibinfo{volume}{78}}, \bibinfo{pages}{031906}
  (\bibinfo{year}{2008}).
  
\bibitem{O4}
\bibinfo{author}{Avelino, P. P.} \emph{et~al.}
\newblock \bibinfo{title}{Junctions and spiral patterns in generalized rock-paper-scissors models}.
\newblock \emph{\bibinfo{journal}{Phys. Rev. E}}
  \textbf{\bibinfo{volume}{86}}, \bibinfo{pages}{036112}
  (\bibinfo{year}{2012}).


\bibitem{O5}
\bibinfo{author}{Szolnoki, A.} \& \bibinfo{author}{Perc, M.}
\newblock \bibinfo{title}{Zealots tame oscillations in the spatial
  rock-paper-scissors game}.
\newblock \emph{\bibinfo{journal}{Phys. Rev. E}}
  \textbf{\bibinfo{volume}{93}}, \bibinfo{pages}{062307}
  (\bibinfo{year}{2016}).


\bibitem{O6}
\bibinfo{author}{Szolnoki, A.} \emph{et~al.}
\newblock \bibinfo{title}{Cyclic dominance in evolutionary games: a review}.
\newblock \emph{\bibinfo{journal}{Journal of The Royal Society Interface}}
  \textbf{\bibinfo{volume}{11}}, \bibinfo{pages}{20140735}
  (\bibinfo{year}{2014}).

\bibitem{O7}
\bibinfo{author}{Szolnoki, A.} \& \bibinfo{author}{Perc, M.}
\newblock \bibinfo{title}{Biodiversity in models of cyclic dominance is
  preserved by heterogeneity in site-specific invasion rates}.
\newblock \emph{\bibinfo{journal}{Scientific Reports}}
  \textbf{\bibinfo{volume}{6}}, \bibinfo{pages}{38608} (\bibinfo{year}{2016}).


\bibitem{R1}
\bibinfo{author}{Juul, J.}, \bibinfo{author}{Sneppen, K.} \&
  \bibinfo{author}{Mathiesen, J.}
\newblock \bibinfo{title}{Labyrinthine clustering in a spatial
  rock-paper-scissors ecosystem}.
\newblock \emph{\bibinfo{journal}{Phys. Rev. E}}
  \textbf{\bibinfo{volume}{87}}, \bibinfo{pages}{042702}
  (\bibinfo{year}{2013}).

\bibitem{R2}
\bibinfo{author}{Vukov, J.}, \bibinfo{author}{Szolnoki, A.} \&
  \bibinfo{author}{Szab{\'{o}}, G.}
\newblock \bibinfo{title}{Diverging fluctuations in a spatial five-species
  cyclic dominance game}.
\newblock \emph{\bibinfo{journal}{Phys. Rev. E}}
  \textbf{\bibinfo{volume}{88}}, \bibinfo{pages}{022123}
  (\bibinfo{year}{2013}).

\bibitem{R3}
\bibinfo{author}{Szolnoki, A.}, \bibinfo{author}{Vukov, J.} \&
  \bibinfo{author}{Perc, M.}
\newblock \bibinfo{title}{From pairwise to group interactions in games of
  cyclic dominance}.
\newblock \emph{\bibinfo{journal}{Phys. Rev. E}}
  \textbf{\bibinfo{volume}{89}}, \bibinfo{pages}{062125}
  (\bibinfo{year}{2014}).

\bibitem{13}
\bibinfo{author}{Hamming, R.~W.}
\newblock \bibinfo{title}{Error detecting and error correcting codes}.
\newblock \emph{\bibinfo{journal}{Bell System Technical Journal}}
  \textbf{\bibinfo{volume}{29}}, \bibinfo{pages}{147--160}
  (\bibinfo{year}{1950}).


\bibitem{14}
\bibinfo{author}{Ramos, J. G. G.~S.}, \bibinfo{author}{Bazeia, D.},
  \bibinfo{author}{Hussein, M.~S.} \& \bibinfo{author}{Lewenkopf, C.~H.}
\newblock \bibinfo{title}{Conductance peaks in open quantum dots}.
\newblock \emph{\bibinfo{journal}{Phys. Rev. Lett.}}
  \textbf{\bibinfo{volume}{107}}, \bibinfo{pages}{176807}
  (\bibinfo{year}{2011}).
  
\bibitem{N1}
\bibinfo{author}{Jiang, L.-L.}, \bibinfo{author}{Zhou, T.},
\bibinfo{author}{Perc, M.}, \&  \bibinfo{author}{Wang, B.-H.}
\newblock \bibinfo{title}{Effects of competition on pattern formation in the rock-paper-scissors game}.
\newblock \emph{\bibinfo{journal}{Phys. Rev. E}}
  \textbf{\bibinfo{volume}{84}}, \bibinfo{pages}{021912}
  (\bibinfo{year}{2011}).

\bibitem{N2}
\bibinfo{author}{Cheng, H., {\it et al}}.
\newblock \bibinfo{title}{Mesoscopic Interactions and Species Coexistence in Evolutionary Game Dynamics of Cyclic Competitions}.
\newblock \emph{\bibinfo{journal}{Scientific Reports}}
  \textbf{\bibinfo{volume}{4}}, \bibinfo{pages}{7486}
  (\bibinfo{year}{2014}).


\bibitem{15}
\bibinfo{author}{R\"{o}ssler, O.}
\newblock \bibinfo{title}{An equation for continuous chaos}.
\newblock \emph{\bibinfo{journal}{Physics Letters A}}
  \textbf{\bibinfo{volume}{57}}, \bibinfo{pages}{397--398}
  (\bibinfo{year}{1976}).



\end{thebibliography}
\end{document}